\documentclass[sigconf]{acmart}
\AtBeginDocument{%
  \providecommand\BibTeX{{%
    Bib\TeX}}}

\microtypesetup{expansion=false}
\usepackage{graphicx}
\usepackage{subfiles}

\usepackage{todonotes}
\usepackage{booktabs}
\usepackage{multirow}
\PassOptionsToPackage{hyphens}{url}\usepackage{hyperref}

\usepackage{amsmath}
\DeclareMathOperator*{\argmax}{arg\,max}
\DeclareMathOperator*{\argmin}{arg\,min}

\usepackage{caption}
\usepackage{subcaption}

\usepackage{xcolor}

\usepackage{pifont}%

\AtBeginDocument{%
  \providecommand\BibTeX{{%
    \normalfont B\kern-0.5em{\scshape i\kern-0.25em b}\kern-0.8em\TeX}}}

\copyrightyear{2026}
\acmYear{2026}
\setcopyright{cc}
\setcctype{by}
\acmConference[SIGIR '26]{Proceedings of the 49th International ACM SIGIR Conference on Research and Development in Information Retrieval}{July 20--24, 2026}{Melbourne, VIC, Australia}
\acmBooktitle{Proceedings of the 49th International ACM SIGIR Conference on Research and Development in Information Retrieval (SIGIR '26), July 20--24, 2026, Melbourne, VIC, Australia}
\acmDOI{10.1145/3805712.3809920}
\acmISBN{979-8-4007-2599-9/2026/07}

\newcommand{\name}{ColBERTSaR}

\begin{document}

\widowpenalties 3 2001 2002 0

\title{\name: Sparsified ColBERT Index via Product Quantization}

\settopmatter{authorsperrow=4}

\author{Eugene Yang}
\affiliation{
  \institution{Johns Hopkins University}
  \city{Baltimore}
  \state{MD}
  \country{USA}
}
\email{eugene.yang@jhu.edu}

\author{Andrew Yates}
\affiliation{
  \institution{Johns Hopkins University}
  \city{Baltimore}
  \state{MD}
  \country{USA}
}
\email{andrew.yates@jhu.edu}

\author{Dawn Lawrie}
\affiliation{
  \institution{Johns Hopkins University}
  \city{Baltimore}
  \state{MD}
  \country{USA}
}
\email{lawrie@jhu.edu}

\author{James Mayfield}
\affiliation{
  \institution{Johns Hopkins University}
  \city{Baltimore}
  \state{MD}
  \country{USA}
}
\email{mayfield@jhu.edu}

\author{Saron Samuel}
\affiliation{
    \institution{Johns Hopkins University}
    \city{Baltimore}
    \state{MD}
    \country{USA}
}
\email{ssamue21@jhu.edu}

\author{Rohan Jha}
\affiliation{
    \institution{Johns Hopkins University}
    \city{Baltimore}
    \state{MD}
    \country{USA}
}
\email{rjha5@jhu.edu}

\renewcommand{\shortauthors}{Yang, et al.}

\begin{abstract}
While ColBERT is an effective neural retrieval architecture,
it requires a heavy index structure to support candidate set retrieval based on approximated token embeddings,
gathering and decompressing document token embeddings,
and applying the MaxSim operation.
Indexes in PLAID and similar ColBERT implementations
require five to ten times the disk storage of the original raw text,
which limits their scalability.
Furthermore, prior work has identified that the gathering and decompression stages
are the primary inefficiencies at query time.
Limiting the number of document tokens that must be gathered by thresholding and score approximation
does not eliminate the need for the entire index to support ad hoc queries.
In this work, we propose an embedding quantization approach that turns a ColBERT index into a true inverted index.
We show that, theoretically, ColBERT with embedding quantization is equivalent to learned-sparse retrieval
except for the scoring mechanism.
Empirically, we demonstrate that our index is 50-70\% smaller than a one-bit PLAID index
while retaining retrieval effectiveness. 
\end{abstract}



\maketitle

\section{Introduction}

Multi-vector dense retrieval models, such as ColBERT~\cite{khattab2020colbert} and XTR~\cite{lee2024rethinking}
are effective because of their expressiveness in modeling. 
While similar to single-vector dense models that represent information with a single contextualized embedding,
such as DPR~\cite{karpukhin2020dense} and Qwen3-Embedding~\cite{zhang2025qwen3embedding},
multi-vector models use multiple vectors for each document to capture more granular information~\cite{formal2021white}. 
However, the number of vectors produced depends on the length of the documents;
this creates efficiency challenges because this number
is typically several orders of magnitude greater than for single vector representations. 
Subsequent work (PLAID~\cite{santhanam2022plaid}) addresses this problem
by storing the vectors in an inverted file index (IVF)
and by using product quantization via heavily compressed residuals~\cite{johnson2019billion, ge2013optimized, santhanam2022plaid}.
Abstractly, at search time, PLAID first gathers document vectors that are similar to the query token vectors from the IVF,
filters them using heuristics,
and finally decompresses all document token vectors of the candidate documents to perform the actual scoring. 
Prior work has explored techniques such as token routing~\cite{li2023citadel},
score imputation~\cite{lee2024rethinking, scheerer2025warp},
fast decompression~\cite{scheerer2025warp}, instruction set optimization~\cite{nardini2024efficient}, 
and replacing the initial candidate search with more efficient models~\cite{macavaney2024reproducibility, formal2024splate}
or even replacing the entire backbone with a smaller model~\cite{wang2023reproducibility} to reduce query latency. 
However, these lines of work do not address the index size problem. 
For one million documents, each with 512 tokens
(a moderate size for a collection taking 1 to 3 GB to store without compression),
PLAID still takes more than 7.8 GB to store aggressive 1-bit compressed residuals for the default ColBERT embedding size of 128. 
This size doubles or quadruples if using 2 or 4-bit compression~\cite{santhanam2022plaid},
becoming an order of magnitude larger than the raw text. 

Sparse retrieval, on the other hand, has a much smaller storage footprint. 
Since each document is represented by its unique tokens and impact scores~\cite{mackenzie2022accelerating},
the index size is much smaller than in ColBERT,
where each token maps to a dense vector
(see Table~\ref{tab:neuclir-index-size} for comparison). 
Recent work in learned sparse retrieval (LSR)
such as SPLADE~\cite{formal2021splade, formal2021spladev2, lassance2024spladev3}
and MILCO~\cite{nguyen2025milco}
has improved model effectiveness substantially
by mapping each document to a set of inferred tokens;
this still keeps the index size manageable~\cite{nguyen2023unified, mallia2021learning}. 

Interestingly, without the residual,
PLAID is essentially a sparse retrieval with a binary document weight
(more in the next section). 
However, residuals, while composing most of the index storage,
are critical for calculating the approximated ColBERT score in PLAID
(see Table~\ref{tab:beir} for empirical results). 
To remove the dependency on residual vectors we introduce \name{},
a sparse approximation of the MaxSim score of any ColBERT-style model
without retraining the retrieval model and storing the residuals. 
\name{} is a drop-in replacement for PLAID that transforms ColBERT into a sparse retrieval model.
\name{} substantially reduces PLAID's index size while providing competitive effectiveness. 
In this work, we 1) provide a theoretical derivation of the \name{} sparse approximation;
2) establish strong connections between ColBERT and LSR models;
and 3) present empirical results on monolingual, cross-language, and multilingual retrieval
using a proof-of-concept implementation.\footnote{\url{https://github.com/hltcoe/ColBERTSaR}}

\section{\name}

\subsection{Sparse Approximation of MaxSim}

MaxSim, proposed by \citet{khattab2020colbert},
aggregates the pairwise similarities between query and document token embeddings,
which are
implemented as the dot products of L2-normalized embeddings. 
Let $q_i\in q$ and $d_j\in d$ be the token embeddings of query $q$ and document $d$.
MaxSim can be written as 
\begin{equation}
    Score(q, d) = \sum_{i=1}^{|q|} \max_{j=1}^{|d|} \left(q_i \cdot d_j\right)
\end{equation}
where $|q|$ and $|d|$ denote the length of the query and document. 

While expressive and effective, this process is expensive in two ways.
First, while all inner product pairs are needed to find the maximum,
only a few contribute to the final score
(specifically $|q|/|d|$,
which is %
6.25\% when query and document lengths are 32 and 512, a common setting).
Second, storing $|d|$ embeddings is expensive.
ColBERT~\cite{khattab2020colbert} stored all embeddings with 16-bit floats for each of the 128 dimensions.
PLAID~\cite{santhanam2022plaid} further compressed the embeddings as a cluster centroid and a residual
with several bits per coordinate.
The MaxSim scoring function can be rewritten as:
\begin{equation}
    Score(q, d) = \sum_{i=1}^{|q|} \max_{j=1}^{|d|} \left( q_i \cdot \left(c_{d_j} + r_j\right) \right)
\end{equation}
where $c_{d_j}$ is the closest anchor to $d_j$ and $r_j=d_j-c_{d_j}$ is the residual embedding
(which can be assigned at indexing time).\footnote{
To avoid defining excessive variables, 
we abuse notation to use $d_j$ in subscript to denote the closest anchor ID,
which formally is $c_{d_j} = \argmax_{c_k} c_k \cdot d_j$.}
Anchor matrix $C$, with $K$ centroids, each as a column vector, $c_k$, 
is fitted using K-means clustering on a sample of the document token embeddings.
MaxSim calculation can be further optimized by only taking the maximum
of document embeddings that are in the $n$ closest partitions of $q_i$ instead of over all anchors $c_k$,
which is the \textit{nprobe} parameter in ANN algorithms~\cite{jayaram2019diskann}.

If the residual norms are small
(generally the assumption when using small nprobe), 
residual vectors can be omitted.
This approximation makes the function depend only on the query/centroid inner products,
which are free since we already calculated them to determine the closest clusters to $q_i$ when using $nprobe$.
The scoring function can then be further simplified as:
\begin{align}
    Score^S(q, d) &= \sum_{i=1}^{|q|} \max_{j=1}^{|d|} \left( q_i \cdot c_{d_j} \right) 
                  = \sum_{i=1} \max_{k\in v_d} q_i \cdot c_k \nonumber \\
                 &= \sum_{i=1}^{|q|} \max_{k=1}^{K} q_i \cdot c_k \cdot \mathbb{1}(k\in v_d)
    \label{eq:centroid-approx}
\end{align}
where $v_d = \{ c_{d_j}| \forall d_j \in d\}$ is the set of anchors that all tokens in document $d$ are closest to
and $\mathbb{1}(\cdot)$ is the indicator function. 
This is also the first stage candidate gathering process in PLAID. 

The core of Equation~\ref{eq:centroid-approx} is a dynamic TF-IDF function. 
Interpreting the anchors $c_k$ as a sparse index vocabulary,
the inner product between each $q_i$ and $c_k$ can be seen as a query token weighting,
or a query-specific inverse document frequency.
This allows us to use the inverted index infrastructure for sparse retrieval.
Using this \textit{residual-free product quantization},
ColBERT is a learned sparse retrieval model with complex score aggregation. 

\subsection{Query-Aware and Unsupervised Anchor Optimization}

Fitting the anchors using K-means clustering
minimizes the distance from each training sample embedding to the centroids;
this can be written as a minimization problem~\cite{armacki2022gradient}.
Assuming $M$ training examples and $K$ centroids, the target can be written:
\begin{equation}
    \min_{C} \sum_{j=1}^M \min_{k=1}^K \left(c_k - x_j\right)^2 \label{eq:kmeans}
\end{equation}
where $x_j$ are the training embeddings.
In other words, K-means minimizes the L2 norm of residual vectors $r$,
which is the difference between its closest centroid and itself.
Optimization usually uses an iterative algorithm such as E-M
but can also be done with gradient descent~\cite{armacki2022gradient};
this gives us more flexibility for extensions. 

However, the approximation error between true MaxSim and the residual-free quantization
depends not only on the size but also on the direction of the residual vectors,
which differs from the K-means clustering optimization goal.
If all residual vectors are zero
there is no approximation error.
Otherwise, let $m(i) = \argmax_j q_i \cdot d_j$,
the approximation error can be written as:
\begin{align*}
    Score(q, d) - Score^S(q, d)  
    &= \sum_i q_i \cdot \left( d_{m(i)} - c_{m(i)} \right) = \sum_i q_i \cdot r_{m(i)} 
\end{align*}
which is the sum of the dot product between each query token embedding
and the residual vector of its matching document token.
Although minimizing the norm of the residual vectors $r$ can indirectly reduce approximation error,
we can directly optimize for the approximation error by modifying Equation~\ref{eq:kmeans} to be query-aware:
\begin{equation}
    \min_{C} \sum_{i=1}^N \sum_{j=1}^M q_i \cdot \argmin_{c_k, \forall k\in[K]} \left(c_k - x_j\right)^2
    \label{eq:query-aware-anchors}
\end{equation}
where $N$ is the number of training queries.
We call this \textbf{query-aware anchor optimization}. 

However, Equation~\ref{eq:query-aware-anchors} requires training queries,
which are usually not present during indexing.
Although they can be mined from existing or publicly available query logs
(feasible in production settings),
the assumed query distribution may be different from actual queries. 
We can, instead, use in-batch training examples as pseudo-queries
to remove the dependency on training queries.
Let $B$ be training mini-batch size.
Mini-batch minimization can be written as:
\begin{equation}
    \min_{C} \sum_{i=1}^B \sum_{j=1}^B x_i \cdot \argmin_{c_k, \forall k\in[K]} \left(c_k - x_j\right)^2
    \label{eq:unsupervised-anchors}
\end{equation}
which forms the \textbf{unsupervised anchor optimization} problem. 

Equation~\ref{eq:unsupervised-anchors} assumes that query embeddings and document token embeddings are similar,
which sometimes is false.
However, it augments training optimization using only document token vectors without additional resources.  
To simulate scenarios where query logs or external queries exist,
we explore using Equation~\ref{eq:query-aware-anchors} with training queries from evaluation collections
and from MS MARCO 
(see Section~\ref{sec:results:sources}).

\subsection{Sparse Indexing and Retrieval}

Indexing begins by sampling document token embeddings
to form a training set for the anchor matrix $C$.
The number of anchors $K$ is a hyperparameter dependent on the number of documents being indexed. 
Anchors created during training are used to sparsify the document token embeddings.
Each document token is assigned to its closest anchor based on the token embedding,
forming a set of documents associated with each anchor point. 
At search time, the $n$ closest anchors to each query token embedding are identified.
All documents in at least one anchor set are considered candidate documents. 
We then use a forward index that maps each document
to its corresponding anchors to produce $Score^S(q, d)$ (Equation~\ref{eq:centroid-approx}). 

\subsubsection{Sparse Indexing}

To reduce disk I/O, the collection is processed in mini-batches and stored in multiple chunks.
After producing ColBERT token embeddings $d_j$ for each document $d$,
we calculate the dot product of $d_j$ with the trained anchor matrix $c$.
The coordinate ID of the highest scoring anchor
is used as the mapped anchor for each document token $d_j$. 
Each chunk is an inverted mapping from an anchor ID to the set of doc IDs containing it. 

After all chunks are stored, we merge them
using $n$-way merge to produce the final inverted index.
For simplicity, we use SciPy CSR Matrix to store the inverted index
and produce the forward index by inverting the inverted matrix. 
While engineering optimizations
such as blocking and sketch vectors can further improve efficiency,
this simple process is a proof-of-concept for storing sparse document representations
to support ColBERT scoring approximation. 

\subsubsection{Searching}

Each query is first encoded using ColBERT to get
query token embeddings $q_i$.
The dot product with the anchor matrix $C$
produces full query token-to-anchor scores $S\in \mathbb{R}^{N\times K}$.

First stage scoring selects the top $n$ anchors 
(i.e., \texttt{nprobe}) for each query token $q_i$. 
For each $q_i$, we collect all document IDs from the postings list across the $n$ anchors
to approximate Equation~\ref{eq:centroid-approx} with only the top-$n$ anchors,
identical to PLAID~\cite{santhanam2022plaid}. 
This can be done efficiently with only a single pass over the inverted index. 

Second stage scoring
uses the forward index to map the top $k$ first-stage documents to the anchor IDs they contain. 
It then slices the query token-anchor score matrix $S$ using these IDs
and uses Equation~\ref{eq:centroid-approx} to produce $Score^S$.

Since the inverted index traversal is not a standard SciPy operation,
we implemented the single-pass traversal algorithm in Cython,
which produced executable Python functions.
Optimization of the max operator in an inverted index requires more investigation, which we leave to future work. 
Therefore, in this work, we omit the latency analysis and focus on the trade-off between index size and retrieval effectiveness.

\section{Experiments}

We evaluate \name{} on BEIR~\cite{thakur2021beir},
NeuCLIRBench~\cite{lawrie2025neuclirbench} (English queries with Chinese, Persian and Russian documents),
and NeuCLIRTech~\cite{lawrie2026neuclirtech} (English queries and Chinese academic abstracts),
which cover monolingual, cross-language (CLIR), and multilingual (MLIR) ad hoc retrieval. 
In the CLIR setting of NeuCLIRBench, we use English queries and retrieve documents from each of the three language-specific collections (Chinese, Persian and Russian). 
In the MLIR setting, queries are still in English, but the document collection is a mixture of the three. 

To demonstrate the space reduction and effectiveness,
we compare \name{} to PLAID using three backbone ColBERT models:
ColBERT-Small-v1~\cite{clavie2024jacolbertv2}\footnote{\url{https://huggingface.co/answerdotai/answerai-colbert-small-v1}} and 
English-Trained PLAID-X~\cite{yang2024translate}\footnote{\url{https://huggingface.co/hltcoe/plaidx-large-eng-tdist-mt5xxl-engeng}} for monolingual English retrieval on BEIR,
and MTD PLAID-X~\cite{yang2024distillation}\footnote{\url{https://huggingface.co/hltcoe/plaidx-large-neuclir-mtd-mix-passages-mt5xxl-engeng}} for CLIR and MLIR on NeuCLIRBench and NeuCLIRTech. 
All documents are split into passages of 512 tokens and aggregate passage scores with MaxP.

As baselines, we compare with BM25 (machine-translated documents, i.e., DT, on NeuCLIRBench and NeuCLIRTech provided by the benchmark to perform lexical matching on English tokens) as the lexical sparse retrieval alternative. 
We also compare with SPLADEv3~\cite{lassance2024spladev3} for BEIR,
and MILCO~\cite{nguyen2025milco}, a state-of-the-art multilingual LSR model, for CLIR and MLIR as the LSR alternatives.

\begin{table}[t]
    \centering
\caption{nDCG@20 on NeuCLIRBench and NeuCLIRTech (Tech). The CLIR column is the average over three languages: Chinese (zho), Persian (fas), and Russian (rus). Since BM25 can only support lexical matching, we translated the documents (DT) into English. }\label{tab:neuclir}
\begin{tabular}{l|ccc|c|c|c}
\toprule
     & \multicolumn{5}{c|}{NeuCLIRBench} & Tech \\
\midrule 
{}   & zho & fas & rus & CLIR  & MLIR  & CLIR \\
\midrule
BM25 w/ DT & 0.439 & 0.447 & 0.400 & 0.429 & 0.349 & 0.234 \\  
MILCO      & 0.431 & 0.494 & 0.476 & 0.467 & 0.395 & 0.264\\
\midrule
PLAID 1bit & 0.495 & 0.529 & 0.463 & 0.495 & 0.396 & 0.362 \\
ColBERTSaR  & 0.471 & 0.529 & 0.475 & 0.492 & 0.385 & 0.348\\
\bottomrule
\end{tabular}

\end{table}

For fitting \name{} anchors, we fix the number of anchors at 500k for NeuCLIRTech and small BEIR datasets
(datasets with fewer than 1M passages; results summarized at the left of Table~\ref{tab:beir}), 
and one million for others. 
The number of sampled training passages follows the sampling rate in PLAID for fitting the K-means clusters,
which is $16\times\sqrt{\bar{|d|}\times D}$ where $\bar{|d|}$ is the assumed document length (120 by default)
and $D$ is the number of documents in the collection. 
We use eight NVIDIA V100 GPUs and a learning rate of $10^{-4}$
with a per-device batch size of 2048 vectors for 100k training steps using fp16.

\begin{table*}[t]
\centering
\caption{nDCG@10 on BEIR datasets using \name{} and PLAID with one bit residual compression with two ColBERT backbone models.
The bottom two rows provide ablations on using different training queries during anchor optimization.}\label{tab:beir}

\resizebox{\linewidth}{!}{
\begin{tabular}{l|ccccccc|ccccccc|c}
\toprule
 & argu & fiqa & nfc & quora & scidoc & scifact & touc & covid & cli. & dbp. & fever & hotpot & msm & nq & Avg \\
\midrule
BM25 & 0.300 & 0.236 & 0.322 & 0.789 & 0.149 & 0.679 & \textbf{0.641} & 0.595 & 0.165 & 0.318 & 0.651 & 0.633 & 0.512 & 0.305 & 0.450 \\
SPLADEv3 & \textbf{0.509} & 0.374 & 0.357 & 0.814 & 0.158 & 0.710 & 0.293 & 0.748 & \underline{0.233} & \textbf{0.450} & 0.796 & \underline{0.692} & 0.440 & \textbf{0.586} & \underline{0.511} \\
\midrule
\multicolumn{16}{l}{ColBERT Model: \texttt{PLAID XLMR ET}} \\
\midrule
PLAID 1bit & 0.324 & \textbf{0.410} & 0.343 & \underline{0.850} & 0.164 & 0.698 & 0.247 & 0.428 & 0.214 & 0.412 & \underline{0.817} & 0.590 & \underline{0.703} & 0.535 & 0.481 \\
\name & 0.339 & 0.376 & 0.347 & 0.798 & 0.164 & 0.671 & 0.326 & 0.550 & 0.147 & 0.334 & 0.674 & 0.481 & 0.565 & 0.456 & 0.445 \\
\midrule
\multicolumn{16}{l}{ColBERT Model: \texttt{colbert-small-v1}} \\
\midrule
PLAID 1bit & 0.314 & \underline{0.398} & 0.353 & \textbf{0.868} & \textbf{0.179} & 0.734 & 0.367 & \textbf{0.835} & \textbf{0.250} & \underline{0.440} & \textbf{0.895} & \textbf{0.746} & \textbf{0.714} & \underline{0.581} & \textbf{0.548} \\
PLAID 0bit & 0.299 & 0.330 & 0.315 & 0.769 & 0.164 & 0.693 & 0.350 & 0.782 & 0.155 & 0.225 & 0.631 & 0.486 & 0.416 & 0.322 & 0.424 \\
\midrule
\name & 0.322 & 0.376 & 0.362 & 0.840 & 0.176 & \underline{0.737} & 0.389 & 0.771 & 0.152 & 0.345 & 0.716 & 0.589 & 0.618 & 0.462 & 0.490 \\
\enspace+ BM25 & \underline{0.343} & 0.338 & 0.355 & 0.844 & 0.169 & 0.730 & \underline{0.560} & 0.717 & 0.177 & 0.366 & 0.741 & 0.670 & 0.612 & 0.439 & 0.505 \\
\enspace w/ Official Train 
 & -- & 0.383 & \textbf{0.367} & -- & -- & 0.735 & -- & -- & -- & -- & 0.721 & 0.593 & 0.632 & -- & -- \\
\enspace w/ MSMARCO 
 & 0.323 & 0.377 & \underline{0.366} & 0.838 & \underline{0.177} & \textbf{0.738} & 0.397 & \underline{0.810} & 0.160 & 0.347 & 0.718 & 0.583 & 0.632 & 0.466 & 0.495 \\
\bottomrule
\end{tabular}
}

\end{table*}

\begin{figure*}[t]
    \centering
    \includegraphics[width=\linewidth]{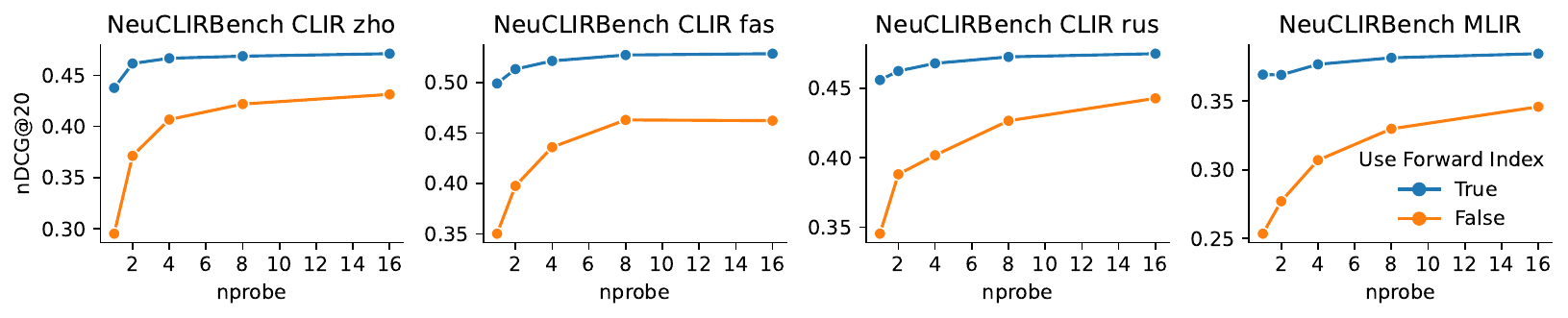}
    \vspace{-2.5em}
    \caption{\name{} nDCG@20 on NeuCLIRBench with different \texttt{nprobe} using  MTD PLAID-X.}
    \label{fig:nprobe}
\end{figure*}

\begin{table}[t]

\caption{Index Size in GB on NeuCLIRBench using different retrieval engines. The top row reports the number of documents in each subset. }\label{tab:neuclir-index-size}
    \centering
\begin{tabular}{l|rrrr}
\toprule
{}          & zho & fas & rus & MLIR \\
\midrule
\textit{\# of documents} & 3.2M & 2.2M & 4.6M & 10.0M \\
\midrule
BM25 w/ DT & 3.44 & 2.33 & 4.39 & 10.14 \\
MILCO      & 6.73 & 4.86 & 7.76 & 19.32 \\
\midrule
PLAID 1bit & 64.53 & 46.50 & 80.39 & 189.12 \\
ColBERTSaR & 14.52 & 11.43 & 37.12 & 89.71 \\
\bottomrule
\end{tabular}
\end{table}

\section{Results}

Our experiments show that our \name{} engine provides competitive effectiveness
compared to the PLAID engine with one bit residual compression
while substantially reducing the index size.
Summarized in Table~\ref{tab:neuclir},
\name{} is capable of representing document token vectors effectively
without being aware that query vectors will be in a very different distribution
(i.e., in a different language). 
\name{} demonstrates competitive effectiveness to MILCO,
the state-of-the-art multilingual LSR model,
indicating that our proposed method is an effective way to sparsify a multi-vector dense retrieval model. 
The gap between PLAID and \name{} on the NeuCLIRTech CLIR task
is slightly bigger than on NeuCLIRBench variants,
indicating that technical terminology in the documents
creates challenges for \name{} when representing the document tokens.

\name{} reduces the index size from 77\% (zho) to 53\% (MLIR),\footnote{Scipy requires int64 for larger indices (rus and fas) to store the document IDs instead of int32 for smaller ones.}
as the index size of each subset in the NeuCLIRBench collection is summarized in Table~\ref{tab:neuclir-index-size}. 
While \name{} indices are still larger than MILCO's,
which are Lucene indices,
engineering tricks such as bit packing can minimize the \name{} indices.

Table \ref{tab:beir} summarizes the nDCG@10 scores on the 13 BEIR subsets with two ColBERT base models. 
\name{} maintains 92\% (0.481 to 0.445) and 89\% (0.548 to 0.490) effectiveness
compared to PLAID with one bit residual compression,
with \texttt{PLAID XLMR ET} and \texttt{colbert-small-v1}, respectively. 
The larger effectiveness gaps appear mostly in question-answering subsets
such as Quora, Fever, Hotpot QA, MSMARCO, and Natural Questions.
These subsets generally contain entities with specific meanings
that are challenging for contextualized token embeddings to capture;
this is a known weakness for ColBERT-style models~\cite{santhanam2022colbertv2}. 
\name{} amplifies such weakness since each document token embedding is further mapped to a centroid
that may be mixed with other meanings for each token. 
By fusing with BM25 (last row in Table~\ref{tab:beir}) using reciprocal rank fusion~\cite{cormack2009reciprocal}
we are able to recover some effectiveness on these QA-style queries while hurting other subsets,
resulting in a slight increase in the average nDCG@10. 
However, with some reasonable prior knowledge in query distribution
it is feasible to pick a suitable stack for the task without much overhead in index size or retrieval,
since the BM25 index is relatively cheap to store and search
(see Table~\ref{tab:neuclir-index-size}). 
Furthermore, compared to PLAID with no residuals (0 bit), essentially \name{} by using K-means centroids as anchors, \name{} optimization substantially improves the anchor when representing document tokens with no residuals.%

\subsection{Query Source for Anchor Optimization}\label{sec:results:sources}

While \name{} uses in-batch documents as pseudo queries when fitting the anchors,
we can use other queries during this process by assuming certain query characteristics. 
At the bottom of Table~\ref{tab:beir},
we observe that using the official BEIR training queries or the MS MARCO training queries
may lead to a slight improvement. 
However, training queries are not always available,
and assuming that queries are similar to MS MARCO-style may also be less optimal
(although many ColBERT models were trained on MS MARCO queries~\cite{yang2024translate, yang2024distillation, khattab2020colbert},
so it may be a fair assumption to reintroduce them at the indexing phase). 
Here, using in-batch document tokens as the query is still likely to be the most robust solution,
as it only requires a sample of documents;
these will always be available at the indexing phase. 

\subsection{Searching with Different \texttt{nprobe}}

One of the critical hyperparameters at query time is \texttt{nprobe},
which is a key parameter in ANN search~\cite{jayaram2019diskann} to control the search space at search time. 
Naturally, the higher \texttt{nprobe}, i.e., the more anchors we explore,
the more accurate the first stage of \name{} retrieval is. 
In Figure~\ref{fig:nprobe}, we observe a diminishing return when increasing \texttt{nprobe}. 
Particularly when there is a second stage that uses a forward index to calculate the actual scores
the final effectiveness saturates at around two to four probes. 
When using only the inverted index, effectiveness saturates on smaller collections
(zho and fas in NeuCLIRBench) 
but continues to improve on larger ones.
This shows that the amount of exploration depends on the size of the collection,
which is expected since there are more points in the index to gather scores. 

\section{Conclusion and Future Work}

This work introduces \name{}, a sparse approximation of ColBERT,
that substantially reduces index size with competitive effectiveness. 
This novel approach provides critical insight into connecting multi-vector dense retrieval and LSR. 
Further engineering, such as document term weightings and index traversal tricks,
will be critical to move beyond this proof-of-concept implementation.

\balance
\bibliographystyle{ACM-Reference-Format}
\bibliography{sample-base}


\begin{thebibliography}{31}

%%% the Web address.

\ifx \showCODEN    \undefined \def \showCODEN     #1{\unskip}     \fi
\ifx \showDOI      \undefined \def \showDOI       #1{#1}\fi
\ifx \showISBNx    \undefined \def \showISBNx     #1{\unskip}     \fi
\ifx \showISBNxiii \undefined \def \showISBNxiii  #1{\unskip}     \fi
\ifx \showISSN     \undefined \def \showISSN      #1{\unskip}     \fi
\ifx \showLCCN     \undefined \def \showLCCN      #1{\unskip}     \fi
\ifx \shownote     \undefined \def \shownote      #1{#1}          \fi
\ifx \showarticletitle \undefined \def \showarticletitle #1{#1}   \fi
\ifx \showURL      \undefined \def \showURL       {\relax}        \fi
% invisible to TeX
\providecommand\bibfield[2]{#2}
\providecommand\bibinfo[2]{#2}
\providecommand\natexlab[1]{#1}
\providecommand\showeprint[2][]{arXiv:#2}

\bibitem[\protect\citeauthoryear{Armacki, Bajovic, Jakovetic, and Kar}{Armacki
  et~al\mbox{.}}{2022}]%
        {armacki2022gradient}
\bibfield{author}{\bibinfo{person}{Aleksandar Armacki},
  \bibinfo{person}{Dragana Bajovic}, \bibinfo{person}{Dusan Jakovetic}, {and}
  \bibinfo{person}{Soummya Kar}.} \bibinfo{year}{2022}\natexlab{}.
\newblock \showarticletitle{Gradient based clustering}. In
  \bibinfo{booktitle}{\emph{International Conference on Machine Learning}}.
  PMLR, \bibinfo{pages}{929--947}.
\newblock


\bibitem[\protect\citeauthoryear{Clavi{\'e}}{Clavi{\'e}}{2024}]%
        {clavie2024jacolbertv2}
\bibfield{author}{\bibinfo{person}{Benjamin Clavi{\'e}}.}
  \bibinfo{year}{2024}\natexlab{}.
\newblock \showarticletitle{JaColBERTv2.5: Optimising Multi-Vector Retrievers
  to Create State-of-the-Art Japanese Retrievers with Constrained Resources}.
\newblock \bibinfo{journal}{\emph{arXiv preprint arXiv:2407.20750}}
  (\bibinfo{year}{2024}).
\newblock


\bibitem[\protect\citeauthoryear{Cormack, Clarke, and Buettcher}{Cormack
  et~al\mbox{.}}{2009}]%
        {cormack2009reciprocal}
\bibfield{author}{\bibinfo{person}{Gordon~V Cormack},
  \bibinfo{person}{Charles~LA Clarke}, {and} \bibinfo{person}{Stefan
  Buettcher}.} \bibinfo{year}{2009}\natexlab{}.
\newblock \showarticletitle{Reciprocal rank fusion outperforms condorcet and
  individual rank learning methods}. In \bibinfo{booktitle}{\emph{Proceedings
  of the 32nd international ACM SIGIR conference on Research and development in
  information retrieval}}. \bibinfo{pages}{758--759}.
\newblock


\bibitem[\protect\citeauthoryear{Formal, Clinchant, D{\'e}jean, and
  Lassance}{Formal et~al\mbox{.}}{2024}]%
        {formal2024splate}
\bibfield{author}{\bibinfo{person}{Thibault Formal},
  \bibinfo{person}{St{\'e}phane Clinchant}, \bibinfo{person}{Herv{\'e}
  D{\'e}jean}, {and} \bibinfo{person}{Carlos Lassance}.}
  \bibinfo{year}{2024}\natexlab{}.
\newblock \showarticletitle{Splate: Sparse late interaction retrieval}. In
  \bibinfo{booktitle}{\emph{Proceedings of the 47th International ACM SIGIR
  Conference on Research and Development in Information Retrieval}}.
  \bibinfo{pages}{2635--2640}.
\newblock


\bibitem[\protect\citeauthoryear{Formal, Lassance, Piwowarski, and
  Clinchant}{Formal et~al\mbox{.}}{2021a}]%
        {formal2021spladev2}
\bibfield{author}{\bibinfo{person}{Thibault Formal}, \bibinfo{person}{Carlos
  Lassance}, \bibinfo{person}{Benjamin Piwowarski}, {and}
  \bibinfo{person}{St{\'e}phane Clinchant}.} \bibinfo{year}{2021}\natexlab{a}.
\newblock \showarticletitle{SPLADE v2: Sparse lexical and expansion model for
  information retrieval}.
\newblock \bibinfo{journal}{\emph{arXiv preprint arXiv:2109.10086}}
  (\bibinfo{year}{2021}).
\newblock


\bibitem[\protect\citeauthoryear{Formal, Piwowarski, and Clinchant}{Formal
  et~al\mbox{.}}{2021b}]%
        {formal2021splade}
\bibfield{author}{\bibinfo{person}{Thibault Formal}, \bibinfo{person}{Benjamin
  Piwowarski}, {and} \bibinfo{person}{St{\'e}phane Clinchant}.}
  \bibinfo{year}{2021}\natexlab{b}.
\newblock \showarticletitle{SPLADE: Sparse lexical and expansion model for
  first stage ranking}. In \bibinfo{booktitle}{\emph{Proceedings of the 44th
  International ACM SIGIR Conference on Research and Development in Information
  Retrieval}}. \bibinfo{pages}{2288--2292}.
\newblock


\bibitem[\protect\citeauthoryear{Formal, Piwowarski, and Clinchant}{Formal
  et~al\mbox{.}}{2021c}]%
        {formal2021white}
\bibfield{author}{\bibinfo{person}{Thibault Formal}, \bibinfo{person}{Benjamin
  Piwowarski}, {and} \bibinfo{person}{St{\'e}phane Clinchant}.}
  \bibinfo{year}{2021}\natexlab{c}.
\newblock \showarticletitle{A white box analysis of ColBERT}. In
  \bibinfo{booktitle}{\emph{European Conference on Information Retrieval}}.
  Springer, \bibinfo{pages}{257--263}.
\newblock


\bibitem[\protect\citeauthoryear{Ge, He, Ke, and Sun}{Ge et~al\mbox{.}}{2013}]%
        {ge2013optimized}
\bibfield{author}{\bibinfo{person}{Tiezheng Ge}, \bibinfo{person}{Kaiming He},
  \bibinfo{person}{Qifa Ke}, {and} \bibinfo{person}{Jian Sun}.}
  \bibinfo{year}{2013}\natexlab{}.
\newblock \showarticletitle{Optimized product quantization}.
\newblock \bibinfo{journal}{\emph{IEEE transactions on pattern analysis and
  machine intelligence}} \bibinfo{volume}{36}, \bibinfo{number}{4}
  (\bibinfo{year}{2013}), \bibinfo{pages}{744--755}.
\newblock


\bibitem[\protect\citeauthoryear{Jayaram~Subramanya, Devvrit, Simhadri,
  Krishnawamy, and Kadekodi}{Jayaram~Subramanya et~al\mbox{.}}{2019}]%
        {jayaram2019diskann}
\bibfield{author}{\bibinfo{person}{Suhas Jayaram~Subramanya},
  \bibinfo{person}{Fnu Devvrit}, \bibinfo{person}{Harsha~Vardhan Simhadri},
  \bibinfo{person}{Ravishankar Krishnawamy}, {and} \bibinfo{person}{Rohan
  Kadekodi}.} \bibinfo{year}{2019}\natexlab{}.
\newblock \showarticletitle{Diskann: Fast accurate billion-point nearest
  neighbor search on a single node}.
\newblock \bibinfo{journal}{\emph{Advances in neural information processing
  Systems}}  \bibinfo{volume}{32} (\bibinfo{year}{2019}).
\newblock


\bibitem[\protect\citeauthoryear{Johnson, Douze, and J{\'e}gou}{Johnson
  et~al\mbox{.}}{2019}]%
        {johnson2019billion}
\bibfield{author}{\bibinfo{person}{Jeff Johnson}, \bibinfo{person}{Matthijs
  Douze}, {and} \bibinfo{person}{Herv{\'e} J{\'e}gou}.}
  \bibinfo{year}{2019}\natexlab{}.
\newblock \showarticletitle{Billion-scale similarity search with GPUs}.
\newblock \bibinfo{journal}{\emph{IEEE Transactions on Big Data}}
  \bibinfo{volume}{7}, \bibinfo{number}{3} (\bibinfo{year}{2019}),
  \bibinfo{pages}{535--547}.
\newblock


\bibitem[\protect\citeauthoryear{Karpukhin, Oguz, Min, Lewis, Wu, Edunov, Chen,
  and Yih}{Karpukhin et~al\mbox{.}}{2020}]%
        {karpukhin2020dense}
\bibfield{author}{\bibinfo{person}{Vladimir Karpukhin}, \bibinfo{person}{Barlas
  Oguz}, \bibinfo{person}{Sewon Min}, \bibinfo{person}{Patrick~SH Lewis},
  \bibinfo{person}{Ledell Wu}, \bibinfo{person}{Sergey Edunov},
  \bibinfo{person}{Danqi Chen}, {and} \bibinfo{person}{Wen-tau Yih}.}
  \bibinfo{year}{2020}\natexlab{}.
\newblock \showarticletitle{Dense Passage Retrieval for Open-Domain Question
  Answering.}. In \bibinfo{booktitle}{\emph{EMNLP (1)}}.
  \bibinfo{pages}{6769--6781}.
\newblock


\bibitem[\protect\citeauthoryear{Khattab and Zaharia}{Khattab and
  Zaharia}{2020}]%
        {khattab2020colbert}
\bibfield{author}{\bibinfo{person}{Omar Khattab} {and} \bibinfo{person}{Matei
  Zaharia}.} \bibinfo{year}{2020}\natexlab{}.
\newblock \showarticletitle{Colbert: Efficient and effective passage search via
  contextualized late interaction over bert}. In
  \bibinfo{booktitle}{\emph{Proceedings of the 43rd International ACM SIGIR
  conference on research and development in Information Retrieval}}.
  \bibinfo{pages}{39--48}.
\newblock


\bibitem[\protect\citeauthoryear{Lassance, D{\'e}jean, Formal, and
  Clinchant}{Lassance et~al\mbox{.}}{2024}]%
        {lassance2024spladev3}
\bibfield{author}{\bibinfo{person}{Carlos Lassance}, \bibinfo{person}{Herv{\'e}
  D{\'e}jean}, \bibinfo{person}{Thibault Formal}, {and}
  \bibinfo{person}{St{\'e}phane Clinchant}.} \bibinfo{year}{2024}\natexlab{}.
\newblock \showarticletitle{SPLADE-v3: New baselines for SPLADE}.
\newblock \bibinfo{journal}{\emph{arXiv preprint arXiv:2403.06789}}
  (\bibinfo{year}{2024}).
\newblock


\bibitem[\protect\citeauthoryear{Lawrie, Mayfield, Yang, Yates, MacAvaney,
  Pradeep, Miller, McNamee, and Soldaini}{Lawrie et~al\mbox{.}}{2026}]%
        {lawrie2026neuclirtech}
\bibfield{author}{\bibinfo{person}{Dawn Lawrie}, \bibinfo{person}{James
  Mayfield}, \bibinfo{person}{Eugene Yang}, \bibinfo{person}{Andrew Yates},
  \bibinfo{person}{Sean MacAvaney}, \bibinfo{person}{Ronak Pradeep},
  \bibinfo{person}{Scott Miller}, \bibinfo{person}{Paul McNamee}, {and}
  \bibinfo{person}{Luca Soldaini}.} \bibinfo{year}{2026}\natexlab{}.
\newblock \showarticletitle{NeuCLIRTech: Chinese Monolingual and Cross-Language
  Information Retrieval Evaluation in a Challenging Domain}.
\newblock \bibinfo{journal}{\emph{arXiv preprint arXiv:2602.05334}}
  (\bibinfo{year}{2026}).
\newblock


\bibitem[\protect\citeauthoryear{Lawrie, Mayfield, Yang, Yates, MacAvaney,
  Pradeep, Miller, McNamee, and Soldani}{Lawrie et~al\mbox{.}}{2025}]%
        {lawrie2025neuclirbench}
\bibfield{author}{\bibinfo{person}{Dawn Lawrie}, \bibinfo{person}{James
  Mayfield}, \bibinfo{person}{Eugene Yang}, \bibinfo{person}{Andrew Yates},
  \bibinfo{person}{Sean MacAvaney}, \bibinfo{person}{Ronak Pradeep},
  \bibinfo{person}{Scott Miller}, \bibinfo{person}{Paul McNamee}, {and}
  \bibinfo{person}{Luca Soldani}.} \bibinfo{year}{2025}\natexlab{}.
\newblock \showarticletitle{NeuCLIRBench: A Modern Evaluation Collection for
  Monolingual, Cross-Language, and Multilingual Information Retrieval}.
\newblock \bibinfo{journal}{\emph{arXiv preprint arXiv:2511.14758}}
  (\bibinfo{year}{2025}).
\newblock


\bibitem[\protect\citeauthoryear{Lee, Dai, Duddu, Lei, Naim, Chang, and
  Zhao}{Lee et~al\mbox{.}}{2024}]%
        {lee2024rethinking}
\bibfield{author}{\bibinfo{person}{Jinhyuk Lee}, \bibinfo{person}{Zhuyun Dai},
  \bibinfo{person}{Sai Meher~Karthik Duddu}, \bibinfo{person}{Tao Lei},
  \bibinfo{person}{Iftekhar Naim}, \bibinfo{person}{Ming-Wei Chang}, {and}
  \bibinfo{person}{Vincent Zhao}.} \bibinfo{year}{2024}\natexlab{}.
\newblock \showarticletitle{Rethinking the role of token retrieval in
  multi-vector retrieval}.
\newblock \bibinfo{journal}{\emph{Advances in Neural Information Processing
  Systems}}  \bibinfo{volume}{36} (\bibinfo{year}{2024}).
\newblock


\bibitem[\protect\citeauthoryear{Li, Lin, Oguz, Ghoshal, Lin, Mehdad, Yih, and
  Chen}{Li et~al\mbox{.}}{2023}]%
        {li2023citadel}
\bibfield{author}{\bibinfo{person}{Minghan Li}, \bibinfo{person}{Sheng-Chieh
  Lin}, \bibinfo{person}{Barlas Oguz}, \bibinfo{person}{Asish Ghoshal},
  \bibinfo{person}{Jimmy Lin}, \bibinfo{person}{Yashar Mehdad},
  \bibinfo{person}{Wen-tau Yih}, {and} \bibinfo{person}{Xilun Chen}.}
  \bibinfo{year}{2023}\natexlab{}.
\newblock \showarticletitle{CITADEL: Conditional token interaction via dynamic
  lexical routing for efficient and effective multi-vector retrieval}. In
  \bibinfo{booktitle}{\emph{Proceedings of the 61st Annual Meeting of the
  Association for Computational Linguistics (Volume 1: Long Papers)}}.
  \bibinfo{pages}{11891--11907}.
\newblock


\bibitem[\protect\citeauthoryear{MacAvaney and Tonellotto}{MacAvaney and
  Tonellotto}{2024}]%
        {macavaney2024reproducibility}
\bibfield{author}{\bibinfo{person}{Sean MacAvaney} {and}
  \bibinfo{person}{Nicola Tonellotto}.} \bibinfo{year}{2024}\natexlab{}.
\newblock \showarticletitle{A reproducibility study of plaid}. In
  \bibinfo{booktitle}{\emph{Proceedings of the 47th International ACM SIGIR
  Conference on Research and Development in Information Retrieval}}.
  \bibinfo{pages}{1411--1419}.
\newblock


\bibitem[\protect\citeauthoryear{Mackenzie, Mallia, Moffat, and
  Petri}{Mackenzie et~al\mbox{.}}{2022}]%
        {mackenzie2022accelerating}
\bibfield{author}{\bibinfo{person}{Joel Mackenzie}, \bibinfo{person}{Antonio
  Mallia}, \bibinfo{person}{Alistair Moffat}, {and} \bibinfo{person}{Matthias
  Petri}.} \bibinfo{year}{2022}\natexlab{}.
\newblock \showarticletitle{Accelerating learned sparse indexes via term impact
  decomposition}. In \bibinfo{booktitle}{\emph{Findings of the Association for
  Computational Linguistics: EMNLP 2022}}. \bibinfo{pages}{2830--2842}.
\newblock


\bibitem[\protect\citeauthoryear{Mallia, Khattab, Suel, and Tonellotto}{Mallia
  et~al\mbox{.}}{2021}]%
        {mallia2021learning}
\bibfield{author}{\bibinfo{person}{Antonio Mallia}, \bibinfo{person}{Omar
  Khattab}, \bibinfo{person}{Torsten Suel}, {and} \bibinfo{person}{Nicola
  Tonellotto}.} \bibinfo{year}{2021}\natexlab{}.
\newblock \showarticletitle{Learning passage impacts for inverted indexes}. In
  \bibinfo{booktitle}{\emph{Proceedings of the 44th International ACM SIGIR
  Conference on Research and Development in Information Retrieval}}.
  \bibinfo{pages}{1723--1727}.
\newblock


\bibitem[\protect\citeauthoryear{Nardini, Rulli, and Venturini}{Nardini
  et~al\mbox{.}}{2024}]%
        {nardini2024efficient}
\bibfield{author}{\bibinfo{person}{Franco~Maria Nardini},
  \bibinfo{person}{Cosimo Rulli}, {and} \bibinfo{person}{Rossano Venturini}.}
  \bibinfo{year}{2024}\natexlab{}.
\newblock \showarticletitle{Efficient multi-vector dense retrieval with bit
  vectors}. In \bibinfo{booktitle}{\emph{European Conference on Information
  Retrieval}}. Springer, \bibinfo{pages}{3--17}.
\newblock


\bibitem[\protect\citeauthoryear{Nguyen, Lei, Ju, Yang, and Yates}{Nguyen
  et~al\mbox{.}}{2025}]%
        {nguyen2025milco}
\bibfield{author}{\bibinfo{person}{Thong Nguyen}, \bibinfo{person}{Yibin Lei},
  \bibinfo{person}{Jia-Huei Ju}, \bibinfo{person}{Eugene Yang}, {and}
  \bibinfo{person}{Andrew Yates}.} \bibinfo{year}{2025}\natexlab{}.
\newblock \showarticletitle{Milco: Learned Sparse Retrieval Across Languages
  via a Multilingual Connector}.
\newblock \bibinfo{journal}{\emph{arXiv preprint arXiv:2510.00671}}
  (\bibinfo{year}{2025}).
\newblock


\bibitem[\protect\citeauthoryear{Nguyen, MacAvaney, and Yates}{Nguyen
  et~al\mbox{.}}{2023}]%
        {nguyen2023unified}
\bibfield{author}{\bibinfo{person}{Thong Nguyen}, \bibinfo{person}{Sean
  MacAvaney}, {and} \bibinfo{person}{Andrew Yates}.}
  \bibinfo{year}{2023}\natexlab{}.
\newblock \showarticletitle{A unified framework for learned sparse retrieval}.
  In \bibinfo{booktitle}{\emph{European Conference on Information Retrieval}}.
  Springer, \bibinfo{pages}{101--116}.
\newblock


\bibitem[\protect\citeauthoryear{Santhanam, Khattab, Potts, and
  Zaharia}{Santhanam et~al\mbox{.}}{2022a}]%
        {santhanam2022plaid}
\bibfield{author}{\bibinfo{person}{Keshav Santhanam}, \bibinfo{person}{Omar
  Khattab}, \bibinfo{person}{Christopher Potts}, {and} \bibinfo{person}{Matei
  Zaharia}.} \bibinfo{year}{2022}\natexlab{a}.
\newblock \showarticletitle{PLAID: an efficient engine for late interaction
  retrieval}. In \bibinfo{booktitle}{\emph{Proceedings of the 31st ACM
  International Conference on Information \& Knowledge Management}}.
  \bibinfo{pages}{1747--1756}.
\newblock


\bibitem[\protect\citeauthoryear{Santhanam, Khattab, Saad-Falcon, Potts, and
  Zaharia}{Santhanam et~al\mbox{.}}{2022b}]%
        {santhanam2022colbertv2}
\bibfield{author}{\bibinfo{person}{Keshav Santhanam}, \bibinfo{person}{Omar
  Khattab}, \bibinfo{person}{Jon Saad-Falcon}, \bibinfo{person}{Christopher
  Potts}, {and} \bibinfo{person}{Matei Zaharia}.}
  \bibinfo{year}{2022}\natexlab{b}.
\newblock \showarticletitle{Colbertv2: Effective and efficient retrieval via
  lightweight late interaction}. In \bibinfo{booktitle}{\emph{Proceedings of
  the 2022 Conference of the North American Chapter of the Association for
  Computational Linguistics: Human Language Technologies}}.
  \bibinfo{pages}{3715--3734}.
\newblock


\bibitem[\protect\citeauthoryear{Scheerer, Zaharia, Potts, Alonso, and
  Khattab}{Scheerer et~al\mbox{.}}{2025}]%
        {scheerer2025warp}
\bibfield{author}{\bibinfo{person}{Jan~Luca Scheerer}, \bibinfo{person}{Matei
  Zaharia}, \bibinfo{person}{Christopher Potts}, \bibinfo{person}{Gustavo
  Alonso}, {and} \bibinfo{person}{Omar Khattab}.}
  \bibinfo{year}{2025}\natexlab{}.
\newblock \showarticletitle{WARP: An efficient engine for multi-vector
  retrieval}. In \bibinfo{booktitle}{\emph{Proceedings of the 48th
  international ACM SIGIR conference on research and development in information
  retrieval}}. \bibinfo{pages}{2504--2512}.
\newblock


\bibitem[\protect\citeauthoryear{Thakur, Reimers, R{\"u}ckl{\'e}, Srivastava,
  and Gurevych}{Thakur et~al\mbox{.}}{2021}]%
        {thakur2021beir}
\bibfield{author}{\bibinfo{person}{Nandan Thakur}, \bibinfo{person}{Nils
  Reimers}, \bibinfo{person}{Andreas R{\"u}ckl{\'e}}, \bibinfo{person}{Abhishek
  Srivastava}, {and} \bibinfo{person}{Iryna Gurevych}.}
  \bibinfo{year}{2021}\natexlab{}.
\newblock \showarticletitle{Beir: A heterogenous benchmark for zero-shot
  evaluation of information retrieval models}.
\newblock \bibinfo{journal}{\emph{arXiv preprint arXiv:2104.08663}}
  (\bibinfo{year}{2021}).
\newblock


\bibitem[\protect\citeauthoryear{Wang, Macdonald, Tonellotto, and Ounis}{Wang
  et~al\mbox{.}}{2023}]%
        {wang2023reproducibility}
\bibfield{author}{\bibinfo{person}{Xiao Wang}, \bibinfo{person}{Craig
  Macdonald}, \bibinfo{person}{Nicola Tonellotto}, {and} \bibinfo{person}{Iadh
  Ounis}.} \bibinfo{year}{2023}\natexlab{}.
\newblock \showarticletitle{Reproducibility, replicability, and insights into
  dense multi-representation retrieval models: from colbert to col}. In
  \bibinfo{booktitle}{\emph{Proceedings of the 46th International ACM SIGIR
  Conference on Research and Development in Information Retrieval}}.
  \bibinfo{pages}{2552--2561}.
\newblock


\bibitem[\protect\citeauthoryear{Yang, Lawrie, and Mayfield}{Yang
  et~al\mbox{.}}{2024a}]%
        {yang2024distillation}
\bibfield{author}{\bibinfo{person}{Eugene Yang}, \bibinfo{person}{Dawn Lawrie},
  {and} \bibinfo{person}{James Mayfield}.} \bibinfo{year}{2024}\natexlab{a}.
\newblock \showarticletitle{Distillation for multilingual information
  retrieval}. In \bibinfo{booktitle}{\emph{Proceedings of the 47th
  International ACM SIGIR Conference on Research and Development in Information
  Retrieval}}. \bibinfo{pages}{2368--2373}.
\newblock


\bibitem[\protect\citeauthoryear{Yang, Lawrie, Mayfield, Oard, and Miller}{Yang
  et~al\mbox{.}}{2024b}]%
        {yang2024translate}
\bibfield{author}{\bibinfo{person}{Eugene Yang}, \bibinfo{person}{Dawn Lawrie},
  \bibinfo{person}{James Mayfield}, \bibinfo{person}{Douglas~W Oard}, {and}
  \bibinfo{person}{Scott Miller}.} \bibinfo{year}{2024}\natexlab{b}.
\newblock \showarticletitle{Translate-distill: learning cross-language dense
  retrieval by translation and distillation}. In
  \bibinfo{booktitle}{\emph{European Conference on Information Retrieval}}.
  Springer, \bibinfo{pages}{50--65}.
\newblock


\bibitem[\protect\citeauthoryear{Zhang, Li, Long, Zhang, Lin, Yang, Xie, Yang,
  Liu, Lin, et~al\mbox{.}}{Zhang et~al\mbox{.}}{2025}]%
        {zhang2025qwen3embedding}
\bibfield{author}{\bibinfo{person}{Yanzhao Zhang}, \bibinfo{person}{Mingxin
  Li}, \bibinfo{person}{Dingkun Long}, \bibinfo{person}{Xin Zhang},
  \bibinfo{person}{Huan Lin}, \bibinfo{person}{Baosong Yang},
  \bibinfo{person}{Pengjun Xie}, \bibinfo{person}{An Yang},
  \bibinfo{person}{Dayiheng Liu}, \bibinfo{person}{Junyang Lin},
  {et~al\mbox{.}}} \bibinfo{year}{2025}\natexlab{}.
\newblock \showarticletitle{Qwen3 Embedding: Advancing Text Embedding and
  Reranking Through Foundation Models}.
\newblock \bibinfo{journal}{\emph{arXiv preprint arXiv:2506.05176}}
  (\bibinfo{year}{2025}).
\newblock


\end{thebibliography}

\end{document}